\documentclass[aps,preprint,floatfix,showpacs]{revtex4-1}
\usepackage[pdftex]{color}
\usepackage{latexsym,amssymb,amsmath,gensymb}
\usepackage{graphicx,dcolumn,bm}
\usepackage{kec,bigints}
\usepackage{mathtools}
\begin{document}

\title{Path integrals for actions that are not quadratic in their time derivatives}

\author{Kevin Cahill}
\email{kevinecahill@gmail.com}
\affiliation{Department of Physics \& Astronomy,
University of New Mexico, Albuquerque, New Mexico 
87131, USA}
\affiliation{School of Computational Sciences,\\
Korea Institute for Advanced Study, Seoul 130-722, Korea}

\date{\today}

\begin {abstract}
The standard way to construct a path integral is to use a Legendre transformation to find the hamiltonian, to repeatedly insert complete sets of states into the time-evolution operator, and then to integrate over the momenta.  This procedure is simple when the action is quadratic in its time derivatives, but in most other cases Legendre's transformation is intractable, and the hamiltonian is unknown.  This paper shows how to construct path integrals when one can't find the hamiltonian because the first time derivatives of the fields occur in ways that make a Legendre transformation intractable; it focuses on scalar fields and does not discuss higher-derivative theories or those in which some fields lack time derivatives.
\end {abstract}

\maketitle

\section {Introduction
\label {Introduction} }
Despite the success of renormalization,
infinities remain a major problem
in quantum field theory, one that
has grown more acute 
as cosmological observations
have confirmed the reality of 
dark energy~\cite{PlanckCosmologicalNA}\@.
For if dark energy is the energy of the vacuum,
then we need to be able to compute 
energies in finite theories. 
The ground-state energy
of a theory with hamiltonian \( H \)
is the big \( \b \) limit of 
\( - d \ln \tr \lt[ \exp( \, - \b H ) \rt] / d\b \)\@.
We can study the ground-state energy
of a theory
if we can write the partition function
\( Z(\b) = \tr \lt[ \exp( \, - \b H ) \rt] \)
as a path integral in euclidian space.
\par
If the action is quadratic
in the first time derivatives of the fields,
then the hamiltonian is 
a simple Legendre transform
of the Lagrange density,
and we can use it
to construct the path integral
in the standard manner~\cite{Weinberg1995IX, *CahillXVI}\@.
But if the action is not quadratic
in the time derivatives of the fields,
then the Legendre transform may be
impossible, the
hamiltonian unknown, 
and the path integral a mystery.
\par
Some quantum field theories 
have finite Green's functions
and finite euclidian action 
densities~\cite{Bender1990, Cahill2013NA, *PhysRevD.88.125014NA},
but they have actions that are not quadratic
in the time derivatives of the fields.
The Nambu-Got{\={o}} action of string theory
is not quadratic in the \(\tau\) derivatives of the 
fields \( X^\mu(\s,\tau) \)\@.
Apart from theories with unbroken
supersymmetry, theories with finite energy densities 
tend to have actions that are not quadratic
in the time derivatives of the fields.
The hamiltonians and path integrals
of these theories
are either complicated or unknown.
This paper shows how to construct path integrals 
when one can't find the hamiltonian because the 
first time derivatives of the fields occur in ways that 
make a Legendre transformation 
intractable; it focuses on scalar fields
and does not discuss higher-derivative
theories~\cite{BenderMannheim2007, *
BenderMannheim2008} 
or those in which some 
fields lack time derivatives
\cite{Dirac1950, *Dirac1958, *Dirac1964}\@.

\section {Lagrangians and hamiltonians
\label {Lagrangians and hamiltonians} }

The lagrangian of a theory tells us about symmetries
and equations of motion,
and the hamiltonian tells us how to construct
path integrals and how to compute 
the time evolution of states and their energies. 
The standard way to construct
a path integral is to use a Legendre transformation
to find the hamiltonian \( H \)
from its lagrangian \( L \) and 
to insert complete sets
of eigenstates of the fields \( \phi_j \) and 
of their conjugate momenta \( \pi_j \)
into the time-evolution operator \( \exp( - i t H ) \)
so as to get 
\begin{equation}
\la \phi'' | e^{-i t H} | \phi' \ra = 
\int \exp \lt\{ \int i \lt[  \dot \phi_j \pi_j 
- H(\phi, \pi) )
\rt]  d^4x \rt\} D\phi D\pi 
\label {phi pi path integral}
\end{equation}
in which the time integral is from
0 to \( t \),
and the letters \( \phi \)
and \( \pi \)
stand for all the fields \( \phi_1,\dots, \phi_n \)
and momenta \( \pi_1, \dots, \pi_n \)
of the action~\cite{Weinberg1995IX, *CahillXVI}.
If one can integrate over the momenta and
does so, then one has the usual expression
\begin{equation}
\la \phi'' | e^{-i t H} | \phi' \ra = 
\int \exp \lt[ \int i \, L(\phi, \dot \phi) \,  d^4x \rt] D\phi .
\label {phi path integral}
\end{equation}
\par
This procedure is straightforward when 
the lagrangian is quadratic
in its time derivatives, 
but in most other theories
the formulas that express
the time derivatives 
in terms of the fields and their momenta
are insoluble and the hamiltonian
is unknown.
Most theories are therefore
inaccessible and unexamined.
\par
The solution to this problem is to let
functional integration perform Legendre's 
transformation (Sec.~\ref {The Legendre transformation})
implicitly.
Delta functionals can
impose the relation between
the time derivatives 
and the fields and their momenta
as illustrated by four examples in
Sec.~\ref {A path-integral Legendre transformation}\@.
The cost is a doubling of the fields 
over which one must integrate
and a determinant that makes a ghostly
appearance when it is positive
(Sec.~\ref {Hidden fermionic variables})\@.
Similar delta functionals work
in euclidian space (Sec.~\ref {Euclidian space})\@.
The Nambu-Goto action of string theory
is not quadratic in its time derivatives; 
its path integral is exhibited 
in Sec.~\ref{Strings}\@.

\section {The Legendre transformation
\label {The Legendre transformation} }

\par
In theories of scalar fields,
the momenta  are
derivatives of the action density
\begin{equation}
\pi_j = \frac{\p  L }{\p \dot \phi_j} .
\label {def of pi}
\end{equation}
If one can invert these equations
and write the time derivatives \( \dot \phi_j \)
of the fields in terms of 
the fields \( \phi_\ell \) 
and their momenta \( \pi_\ell \),
then the energy density is
\begin{equation}
H = \sum_{j=1}^n \pi_j \dot \phi_j(\phi,\pi) 
- L (\phi,\dot \phi(\phi,\pi) ) .
\label {energy density}
\end{equation}
\par
When the action is quadratic 
in all the time derivatives,
this Legendre transformation
is easy to do, but in most other cases
no solution is known
even in the absence of constraints.

\section {A path-integral Legendre transformation
\label {A path-integral Legendre transformation} }

Let the field \( \dot \chi(\phi, \pi) \)
be the function of the fields \( \phi_j \) and
their conjugate momenta \( \pi_j \) that
satisfies Legendre's relation
\begin{equation}
\pi_j = \frac{\p  L(\phi,\dot \chi) }{\p \dot \chi_j} .
\label {def of psi}
\end{equation} 
In terms of this in-general-inaccessible function
\( \dot \chi(\phi, \pi) \), the energy density is
\begin{equation}
H(\phi,\pi) = \sum_{j=1}^n \pi_j \dot \chi_j(\phi,\pi) 
- L (\phi,\dot \chi(\phi,\pi) ) 
\label {psi energy density}
\end{equation}
and the path integral 
(\ref {phi pi path integral}) is
\begin{equation}
\la \phi'' | e^{-i t H} | \phi' \ra = {}
\int \exp \lt\{ \int i \lt[  \dot \phi_j \pi_j 
- \lt( \pi_k \dot \chi_k(\phi,\pi) - L(\phi, \dot \chi(\phi,\pi)) \rt)
\rt] d^4x \rt\} D\phi D\pi .
\label {standard path integral in terms of chi}
\end{equation}
Although we don't know what \( \dot \chi(\phi, \pi) \) is,
we still can write this path integral in terms
of a delta functional as
\begin{equation}
   \begin{split}
\la \phi'' | e^{-i t H} | \phi' \ra = {}&
\int \exp \lt\{ \int i \lt[  \dot \phi_j \pi_j 
- \lt( \pi_k \dot \chi_k(\phi,\pi) - L(\phi, \dot \chi(\phi,\pi)) \rt)
\rt] d^4x \rt\} \\
{}& \times \prod_{\ell,x} \lt[ \delta \lt( \dot \chi_\ell(\phi,\pi) 
-  \dot \psi_\ell  \rt) \rt]
D\phi D\pi D \dot \psi .
\label {standard path integral in terms of chi and dot psi}
   \end{split}
\end{equation}
We can use the delta functional to replace
\( \dot \chi \) by \( \dot \psi \) everywhere else:
\begin{equation}
   \begin{split}
\la \phi'' | e^{-i t H} | \phi' \ra = {}&
\int \exp \lt\{ \int i \lt[  \dot \phi_j \pi_j 
- \lt( \pi_k \dot \psi_k(\phi,\pi) - L(\phi, \dot \psi(\phi,\pi)) \rt)
\rt] d^4x \rt\} \\
{}& \times \prod_{\ell, x} 
\lt[ \delta \lt( \dot \chi_\ell(\phi,\pi) 
-  \dot \psi_\ell  \rt) \rt]
D\phi D\pi D \dot \psi .
\label {standard path integral mostly in terms of dot psi}
   \end{split}
\end{equation}
The delta functional \( \d( \dot \chi - \dot \psi ) \)
imposes the relation (\ref {def of psi})
among the fields \( \phi \),
\( \pi \), and \( \dot \psi \)\@.
We now use the delta-function identity
\begin{equation}
   \begin{split}
\prod_{\ell, x} 
\lt[ \delta \lt( \dot \chi_\ell(\phi,\pi) 
-  \dot \psi_\ell  \rt) \rt] ={}&
\prod_{\ell, x} \lt[ \delta \lt( \pi_\ell 
- \frac{ \p  L(\phi, \dot \psi) }{ \p \dot \psi_\ell }  \rt) \rt]
\lt| \det \lt( \frac{ \p \pi_k }
 { \p \dot \psi_\ell  }  \rt) \rt| \\
={}&
\prod_{\ell, x} \lt[ \delta \lt( \pi_\ell 
- \frac{ \p  L(\phi, \dot \psi) }{ \p \dot \psi_\ell }  \rt) \rt]
\lt| \det \lt( \frac{ \p^2  L(\phi, \dot \psi) }
 { \p \dot \psi_k \p \dot \psi_\ell }  \rt) \rt| 
 \label {delta-function identity}
    \end{split}
\end{equation}
to change variables in the delta functional 
introducing the appropriate jacobian
\begin{equation}
   \begin{split}
\la \phi'' | e^{-i t H} | \phi' \ra = {}&
\int \exp \lt\{ \int i \lt[  \dot \phi_j \pi_j 
- \lt( \pi_k \dot \psi_k - L(\phi, \dot \psi) \rt)
\rt] d^4x \rt\} \\
{}& \times  \prod_{\ell, x} \lt[ \delta \lt( \pi_\ell 
- \frac{ \p  L(\phi, \dot \psi) }{ \p \dot \psi_\ell }  \rt) \rt]
\lt| \det \lt( \frac{ \p^2  L(\phi, \dot \psi) }
 { \p \dot \psi_k \p \dot \psi_\ell }  \rt) \rt|
D\phi D\pi D\dot\psi .
\label {cubic path integral}
   \end{split}
\end{equation}
The repeated indices \( j, k, \ell \) are summed
from 1 to \( n \)\@.
The integration is over all fields 
that go from \( \phi' \) to \(  \phi'' \)
in time \( t \) and over all \( \pi \) and \( \dot \psi \)
in that time interval.
Integrating over \( \pi \), we find
\begin{equation}
   \begin{split}
\la \phi'' | e^{-i t H} | \phi' \ra = {}&
\int \exp \lt[ \int i \,  ( \dot \phi_\ell
- \dot \psi_\ell ) \frac{ \p  L(\phi, \dot \psi) }{ \p \dot \psi_\ell }
+ i \, L(\phi, \dot \psi) \,
d^4x \rt] 
\lt| \det \lt( \frac{ \p^2  L(\phi, \dot \psi) }
{ \p \dot \psi_k \p \dot \psi_\ell }  \rt) \rt|
 D\phi D\dot\psi .
\label {the path integral}
   \end{split}
\end{equation}
This functional integral generalizes 
the path integral to theories of scalar fields
in which the action is not quadratic in the
time derivatives of the fields.
A similar formula should work in
theories of vector and tensor fields,
apart from the issue of constraints.
\par
Our first example is a free scalar field
with action density
\begin{equation}
L = \half \, \p_\mu \phi \, \p^\mu \phi 
- \half m^2 \phi^2
\label {free L}
\end{equation}
and canonical momentum
\begin{equation}
\pi = \frac{ \p  L }{ \p \dot \phi }
= \dot \phi .
\end{equation}
The determinant is unity, and
 the proposed path integral
(\ref {the path integral}) is
\begin{equation}
   \begin{split}
\la \phi'' | e^{-i t H} | \phi' \ra = {}&
\int \exp \lt\{ \int i \lt[  L(\phi, \dot \psi) 
+ \dot \psi ( \dot \phi - \dot \psi )
\rt] d^4x \rt\}  D\phi D\dot \psi \\
= {}&
\int \exp \lt\{ \int i \lt[  \half \dot \psi^2 - \half (\grad \phi)^2 
- \half m^2 \phi^2 
+ \dot \psi ( \dot \phi - \dot \psi )
\rt] d^4x \rt\}  D\phi D\dot \psi .
\label {free path integral}
   \end{split}
\end{equation}
Shifting \( \dot \psi \) to  \( \dot \psi + \dot \phi \)
and integrating over \( \dot \psi \), we get
the usual result
\begin{equation}
   \begin{split}
\la \phi'' | e^{-i t H} | \phi' \ra 
= {}&
\int \exp \lt\{ \int i \lt[  \half (\dot \psi + \dot \phi)^2 
- \half (\grad \phi)^2 
- \half m^2 \phi^2 
- (\dot \psi + \dot \phi) \dot \psi )
\rt] d^4x \rt\}  D\phi D\dot \psi \\
= {}&
\int \exp \lt\{ \int i \lt[  \half \dot \phi^2 
- \half (\grad \phi)^2 
- \half m^2 \phi^2 
- \half \dot \psi^2 
\rt] d^4x \rt\}  D\phi D\dot \psi \\
= {}&
\int \exp \lt\{ \int i \lt[  \half \dot \phi^2 
- \half (\grad \phi)^2 
- \half m^2 \phi^2 
\rt] d^4x \rt\}  D\phi \\
{}& = \int \exp 
\lt[ i \int L(\phi, \dot \phi) \, d^4 x \rt] D\phi .
\label {free path integral}
   \end{split}
\end{equation}

\par
Our second example
is a field theory in one dimension, time.
The lagrangian for a relativistic particle
of mass \( m \) 
\begin{equation}
L = - m \sqrt{1 - \dot q^2}
\end{equation}
is not quadratic the
first time derivative \( \dot q \)\@.
This theory is special in that
we can invert the definition
of the momentum
\begin{equation}
p = \frac{\p L}{\p \dot q} = 
\frac{m \dot q}{\sqrt{1 - \dot q^2}} ,
\end{equation}
express \( \dot q \) in terms of \( p \)
\begin{equation}
\dot q = \frac{p}{\sqrt{p^2 + m^2}} ,
\label {dot q}
\end{equation}
and find the hamiltonian 
\begin{equation}
H = \sqrt{p^2 + m^2} .
\label {H}
\end{equation}
The standard double path integral (\ref {phi pi path integral})
then is
\begin{equation}
\la q''|e^{-itH} | q' \ra =
\int \exp\lt[ \int i 
\lt( \dot q p - \sqrt{p^2 + m^2} \rt) dt
\rt] Dq Dp .
\label {double path integral}
\end{equation}
\par
Suppose, however, that 
we were unable to perform
Legendre's transformation
and get these equations 
(\ref {dot q}--\ref {double path integral})\@.
In that case, we could use
the proposed path integral
(\ref {the path integral})
\begin{equation}
\la q''|e^{-itH} | q' \ra =
\int \exp\lt[ \int i 
\lt( \dot q - \dot s \rt) \frac{\p L(q,\dot s)}{\p \dot s} 
+ i L(q,\dot s) \rt] \lt| \frac{\p^2 L(q,\dot s)}
{\p \dot s^2} \rt| Dq D\dot s .
\label {my path integral}
\end{equation}
We then would write
\begin{equation}
   \begin{split}
\la q''|e^{-itH} | q' \ra = {}&
\int \exp\lt[ \int i 
\lt( \dot q - \dot s \rt) \frac{m \dot s}{\sqrt{1 - \dot s^2}}
- i m \sqrt{1 - \dot s^2} \rt] 
\frac{m}{( 1 - \dot s^2 )^{3/2}} Dq D\dot s .   
\label {this path integral}
   \end{split}
\end{equation}
\par
To check that this path integral (\ref {this path integral})
is the same as the standard double path integral 
(\ref {double path integral}), we
change variables in it (\ref {this path integral}), setting 
\begin{equation}
\dot s = \frac {p}{\sqrt{p^2 + m^2}}
\end{equation}
so that
\begin{equation}
d \dot s = \frac{m^2}{(p^2+m^2)^{3/2}} dp .
\end{equation}
We then find that
\begin{equation}
\frac{m}{( 1 - \dot s^2 )^{3/2}} d \dot s = m
\lt(\frac{p^2+m^2}{m^2}\rt)^{3/2}
\frac{m^2}{(p^2+m^2)^{3/2}} dp = dp ,
\end{equation}
and that
\begin{equation}
   \begin{split}
\lt( \dot q - \dot s \rt) \frac{m \dot s}{\sqrt{1 - \dot s^2}}
- m \sqrt{1 - \dot s^2}  ={}&
\dot q p - \frac{p^2}{\sqrt{p^2 + m^2}} 
- \frac{m^2}{\sqrt{p^2 + m^2}} \\
={}& \dot q p  - \sqrt{p^2 + m^2} .
      \end{split}
\end{equation}
Thus the proposed path integral (\ref {the path integral})
reduces in this quantum-mechanical example
to a path integral (\ref {my path integral})
that is the same as the standard
double path integral (\ref {double path integral})
for this example.

\par
Our third example is the scalar
Born-Infeld theory with 
action density
\begin{equation}
L = {} - M^4 
\sqrt{ 1-  M^{-4}
\lt(\dot \phi^2 
- (\grad \phi)^2 - m^2 \phi^2 \rt) }  
\label {Lsqrt}
\end{equation}
which is the field-theory version
of the second example.
The proposed path integral 
(\ref {the path integral}) is
\begin{equation}
   \begin{split}
\la \phi'' | e^{-i t H} | \phi' \ra 
= {}&   \int \exp \lt\{ \int i \lt[  ( \dot \phi
- \dot \psi ) \frac{ \p  L(\phi, \dot \psi)  }{ \p \dot \psi }
+ L(\phi, \dot \psi) 
\rt] d^4x \rt\}  
\lt| \frac{ \p^2  L(\phi, \dot \psi) }{ \p \dot \psi^2  }  \rt|
 D\phi D\dot\psi
 \label {the path integral 3}
   \end{split}
\end{equation}
in which the partial derivatives are
\begin{align} 
\pi(\phi,\dot \psi) = {}&
\frac{ \p  L(\phi, \dot \psi)  }{ \p \dot \psi } = 
\frac{\dot \psi}
{\sqrt{ 1-  M^{-4} \lt(\dot \psi^2 
- (\grad \psi)^2 - m^2 \phi^2 \rt) }} 
\label {equations to invert}
\end{align}
and
\begin{align}
\frac{ \p \pi(\phi,\dot \psi) }{ \p \dot \psi} = {}&
\frac{ \p^2  L(\phi, \dot \psi) }{ \p \dot \psi^2  }
=  \frac{1 + M^{-4} \lt(
(\grad \psi)^2 + m^2 \phi^2 \rt) }
{\sqrt{ 1-  M^{-4} \lt(\dot \psi^2 
- (\grad \psi)^2 - m^2 \phi^2 \rt) }}  .
\label {dpi}
\end{align}
Substituting these formulas into
the proposed path integral
(\ref {the path integral 3}) gives
\begin{equation}
   \begin{split}
\la \phi'' | e^{-i t H} | \phi' \ra 
= {}&   \int \exp \lt\{ \int i \lt[  \lt( \dot \phi
- \dot \psi \rt) \pi (\phi,\dot \psi)
+ L(\phi, \dot \psi) 
\rt] d^4x \rt\}  
\frac{ \p  \pi (\phi,\dot \psi)}{ \p \dot \psi  }  
 D\phi D\dot\psi . 
 \label {the path integral 4}
   \end{split}
\end{equation}
This theory is one of the few 
in which we can solve (\ref {equations to invert})
for the time derivative \( \dot \psi \)
\begin{equation}
\dot \psi = \frac{\pi}{\sqrt{1 + M^{-4} \, \pi^2}}
\sqrt{1 + M^{-4} 
\lt( (\grad \phi)^2 + m^2 \phi^2 \rt)}  
\label {dot psi}
\end{equation}
and find as the hamiltonian density
\begin{equation}
   \begin{split}
H (\phi,\pi (\phi,\dot \psi)) = {}& \pi(\phi,\dot \psi) \dot \psi 
- L(\phi, \dot \psi) \\
= {}& \sqrt{\lt(M^4 + \pi^2\rt)
\lt(M^4 + (\grad \phi)^2 + m^2 \phi^2\rt)}  .
\label {H1}
   \end{split}
\end{equation}
Thus the proposed path integral 
(\ref {the path integral 4}) is
the standard formula
(\ref {phi pi path integral})
\begin{equation}
   \begin{split}
\la \phi'' | e^{-i t H} | \phi' \ra = {}&
\int \exp \lt\{ \int i \lt[ \dot \phi \pi (\phi,\dot \psi) {} 
-  H(\phi,\pi (\phi,\dot \psi)) \rt] d^4x \rt\} 
D\phi D\dot \psi \frac{ \p  \pi }{ \p \dot \psi  } \\
= {}&
\int \exp \lt\{ \int i \lt[ \dot \phi \pi  {} 
-  H(\phi,\pi ) \rt] d^4x \rt\} 
D\phi D\pi \\
= {}&
\int \exp \lt\{ \int i \lt[ \dot \phi \pi  {} 
- \sqrt{\lt(M^4 + \pi^2\rt)\lt(M^4 
+ (\grad \phi)^2 + m^2 \phi^2\rt)} 
\rt] d^4x \rt\} D\phi D\pi .
\label {Born-Infeld path integral}
   \end{split}
\end{equation}
\par
Our fourth example is the theory
defined by the action density
\begin{equation}
L = M^4 \exp( L_0/M^4 )
\end{equation}
in which \( L_0 \) is the action
density (\ref {free L}) of the free field.
The derivatives
of \( L \) are
\begin{equation}
\frac{\p L}{\p \dot \psi} = M^{-4} \dot \psi \, L
\quad \mbox{and} \quad
\frac{\p^2 L}{\p \dot \psi^2}
= M^{-4} ( 1 + M^{-4} \dot \psi^2 ) \, L .
\end{equation}
So the proposed path integral is
\begin{equation}
\begin{split}
\la \phi'' | e^{-i t H} | \phi' \ra = {}&
\int \exp \lt\{ \int i  \lt[ L(\phi, \dot \psi) 
+ \frac{ \p  L(\phi, \dot \psi) }{ \p \dot \psi } 
( \dot \phi - \dot \psi )
\rt] d^4x \rt\} 
\lt| \frac{\p^2 L(\phi, \dot \psi) }{\p \dot \psi^2} \rt|
D\phi D\dot\psi  \\
={}& \int \exp \lt\{ \int i  \lt[ 1
+ \frac{ \dot \psi 
( \dot \phi - \dot \psi ) }
{M^4}
\rt] L(\phi, \dot \psi) \, d^4x \rt\} 
\frac{( 1 + M^{-4} \dot \psi^2 ) L(\phi, \dot \psi)}
{M^4} \,
D\phi D\dot\psi .
\label {basic path integral pi a}
\end{split}
\end{equation}

\section {Hidden fermionic variables
\label {Hidden fermionic variables} }

A determinant is a gaussian integral
\beq
\int e^{-\thet^\dag A \thet} \, 
\prod_{k=1}^n d\thet^*_k d\thet_k 
= \det A 
\label {det is fermionic path integral}
\eeq
as is well-known~\cite{Weinberg1995IX, *CahillXVI}\@.
When the determinant is positive,
we can drop the absolute-value signs
and write the proposed
path integral (\ref {the path integral}) as
\begin{equation}
   \begin{split}
\la \phi'' | e^{-i t H} | \phi' \ra = {}&
\int \exp \lt\{ \int i \lt[  ( \dot \phi_\ell
- \dot \psi_\ell ) \frac{ \p  L(\phi, \dot \psi) }{ \p \dot \psi_\ell }
+ L(\phi, \dot \psi) 
\rt] d^4x \rt\}  
\det \lt( \frac{ \p^2  L }{ \p \dot \psi_k \p \dot \psi_\ell }  \rt)
 D\phi D\dot\psi \\
= {}&
\int \exp \lt\{ \int i \,  ( \dot \phi_\ell
- \dot \psi_\ell ) \frac{ \p  L }{ \p \dot \psi_\ell }
+ i \, L {} - 
\bar \chi_k \frac{ \p^2  L }{ \p \dot \psi_k \p \dot \psi_\ell }
\chi_\ell
\, d^4x \rt\}  
 D\phi D\dot\psi D\bar \chi D\chi
\label {the path integral with chi's}
   \end{split}
\end{equation}
in which \( \phi_\ell \) and \( \psi_\ell \)
are boson fields, and \( \chi_k  \)
are scalar Grassmann fields or ghosts.
\section {Euclidian space
\label {Euclidian space}}

It is easier to evaluate 
path integrals in euclidian space
where in the proposed path integral
(\ref {the path integral}) the integral over \( x^4 \)
runs from 0 to the inverse temperature
\( \b = 1/kT \)
\begin{equation}
   \begin{split}
\la \phi'' | e^{- \b H} | \phi' \ra = {}&
\int \exp \lt[ \int  
\lt( i \dot \phi_j - \dot \psi_j \rt) 
\frac{ \p  L(\phi, \dot \psi) }{ \p \dot \psi_j } 
 + L(\phi, \dot \psi) d^4x \rt]
\lt|  \det \lt( \frac{ \p^2  L (\phi, \dot \psi)}  
 { \p \dot \psi_k \p \dot \psi_\ell }  \rt) \rt|
D\phi D\dot\psi .
\label {basic path integral in euclidian space}
   \end{split}
\end{equation}
\par
In this theory, the mean value of an observable
\( A[\phi] \) in a system at maximum entropy
and inverse temperature \( \b \) is
\begin{equation}
   \begin{split}    
\la A[\phi] \ra ={}& \frac{ \tr \, A[\phi] e^{- \b H} }{ \tr \, e^{- \b H} } 
=
\int A[\phi] \exp \lt[ \int    \lt( i \, \dot \phi_j 
-  \dot \psi_j \rt) \frac{ \p  L }{ \p \dot \psi_j }  
+ L \, d^4x \rt] 
\lt| \det \lt( \frac{ \p^2  L }
 { \p \dot \psi_k \p \dot \psi_\ell }  \rt) \rt|
D\phi D\dot\psi  
\\
{}& \lt/ \int \exp \lt[ \int  \lt( i \, \dot \phi_j 
-  \dot \psi_j \rt) \frac{ \p  L }{ \p \dot \psi_j }  
+ L \, d^4x \rt] 
\lt| \det \lt( \frac{ \p^2  L }
 { \p \dot \psi_k \p \dot \psi_\ell }  \rt) \rt|
 \, D\phi D\dot\psi  \rt. .
\end{split}
\end{equation}
This ratio of path integrals is a ratio 
of mean values
\begin{equation}
\begin{split}    
\la A[\phi] \ra ={}& \lt\la A[\phi]
\exp \lt[ \int  i \, \dot \phi_j \,
 \frac{ \p  L }{ \p \dot \psi_j }  \, d^4x \rt]
\rt\ra
 \lt / 
\lt\la 
\exp \lt[ \int   i \, \dot \phi_j 
\, \frac{ \p  L }{ \p \dot \psi_j } \, d^4x \rt]
\rt\ra \rt.
\end{split}
\end{equation}
each estimated 
by Monte Carlo simulation~\cite{CahillXIV}
in the normalized probability distribution
\begin{equation}
   \begin{split}
   P(\phi, \dot \psi) ={}&
\exp \lt[ \int  \lt( L(\phi, \dot \psi) 
-  \dot \psi_j  \, \frac{ \p  L }{ \p \dot \psi_j }  \rt)
 d^4x \rt] 
\lt| \det \lt( \frac{ \p^2  L }
 { \p \dot \psi_k \p \dot \psi_\ell }  \rt) \rt|
\\
{}& \lt / \int
 \exp \lt[ \int   \lt( L(\phi, \dot \psi) 
 -  \dot \psi_j \, \frac{ \p  L }{ \p \dot \psi_j }  \rt)
  d^4x \rt] 
\lt| \det \lt( \frac{ \p^2  L }
 { \p \dot \psi_k \p \dot \psi_\ell }  \rt) \rt|
D\phi D\dot\psi  \rt. .
   \end{split}
\end{equation}

\section {Strings
\label {Strings} }

The tau or time derivatives
of the coordinate fields \( X^\mu \)
in the Nambu-Got{\={o}} Lagrange density
\beq
L = - \frac{T_0}{c}
\int_0^{\sigma_1}
\sqrt{ \lt( \dot X \cdot X' \rt)^2
- \lt( \dot X \rt)^2 \lt( X' \rt)^2 }
\label {Nambu Goto L}
\eeq
do not occur 
quadratically.
The momenta are
\beq
\mathcal{P}^\tau_\mu = 
\frac{\partial L}
{\partial \dot X^\mu} =
- \frac{T_0}{c} 
\frac{(\dot X \cdot X') X'_\mu 
- (X')^2 \dot X_\mu}
{\sqrt{ \lt( \dot X \cdot X' \rt)^2
- \lt( \dot X \rt)^2 \lt( X' \rt)^2 }}
\label {Ptaumu}
\eeq
and the second derivatives 
of the Lagrange density are~\cite{CahillXIX}
\begin{equation}
   \begin{split}
\frac{\p^2 L}{\p \dot X^\mu \p \dot X^\nu}
= {}& \frac{T_0}{c} \lt[
\frac{\d_{\mu \nu} X'^2 - X'_\mu X'_\nu}
{ \sqrt{ \lt( \dot X \cdot X' \rt)^2
- \lt( \dot X \rt)^2 \lt( X' \rt)^2 } } \rt. \\
{}& \lt.
- \frac{ \lt( (\dot X \cdot X') X'_\mu 
- (X')^2 \dot X_\mu \rt)
\lt( (\dot X \cdot X') X'_\nu 
- (X')^2 \dot X_\nu \rt) }
{ \lt[ \lt( \dot X \cdot X' \rt)^2
- \lt( \dot X \rt)^2 \lt( X' \rt)^2 \rt]^{3/2} }
\rt]  .
\label {NG 2d derivatives}
   \end{split}
\end{equation}
The proposed path integral 
(\ref {the path integral}) for
the Nambu-Goto action is then
\begin{equation}
   \begin{split}
\la X'' | e^{-i \tau H} | X' \ra = {}&
\int  \exp  \lt[ \int i ( \dot X^\mu
- \dot Y^\mu ) \frac{ \p  L(X, \dot Y) }
{ \p \dot Y^\mu }
+ i L(X, \dot Y) 
\, d^4x \rt]  \\
{}& \times \lt| \det  \lt[ \frac{ \p^2  L(X, \dot Y) }
{ \p \dot Y^\mu \p \dot Y^\nu }  \rt] \rt| 
 DX D\dot Y 
\label {the path integral with chi's for NG}
   \end{split}
\end{equation}
in which the formulas (\ref {Ptaumu})
and (\ref {NG 2d derivatives}) 
(with \( \dot X^\mu \to \dot Y^\mu \)) are to be
substituted for the first and second
derivatives of the action density \( L \)
with respect to the tau derivatives
\( \dot Y^\mu \)\@.

\begin{acknowledgments}
Conversations with David Amdahl, Michael Grady,
Sudhakar Prasad, and Shashank Shalgar
advanced this work which arose out of
numerical computations done with David Amdahl at 
the Korea Institute for Advanced Study
and at the  National Energy Research Scientific Computing Center, a DOE Office of Science User Facility supported by the Office of Science of the U.S. Department of Energy under Contract No. DE-AC02-05CH11231.
\end{acknowledgments}
\bibliography{physics}

\end{document}